\documentclass[letter]{aa}
\usepackage{txfonts}
\usepackage{graphicx}
\usepackage{natbib}
\bibpunct{(}{)}{;}{a}{}{,} 
\usepackage{amstext}

\begin{document}

\title{A triple system with a nonaccreting black hole in the inner binary:  A naked-eye analog of LB-1?
\thanks{The authors dedicate this paper to the memory of Stan {\v S}tefl (1955-2014) in sadness and grateful appreciation of his never-tiring alertness that also triggered this work.}
\thanks{Based partly on observations collected at the
    European Southern Observatory, Chile (Prop.\ Nos. 63.H-0080 and 073.D-0274) 
    }
}
%
%
%
   \author{
        Th.~Rivinius \inst{1}
            \and
        D.~Baade\inst{2}
             \and
        P.~Hadrava\inst{3}
            \and
        M.~Heida\inst{2}
            \and
        R.~Klement\inst{4}
}
\offprints{Th.\,Rivinius \\ \email{triviniu@eso.org}}
\institute{
           European Organisation for Astronomical Research in the  Southern Hemisphere (ESO), Casilla 19001, Santiago 19, Chile
           \and
           European Organisation for Astronomical Research in the Southern Hemisphere (ESO), Karl-Schwarzschild-Str.\,2, 85748 Garching b.\,M\"unchen, Germany
           \and
           Astronomical Institute, Academy of Sciences of the Czech Republic, Bo\v{c}n\'{i} II 1401, CZ 141 31 Praha 4, Czech Republic
           \and
           The CHARA Array of Georgia State University, Mount Wilson Observatory, Mount Wilson, CA 91023, USA
}
\date{Received: $<$date$>$; accepted: $<$date$>$}
\authorrunning{Rivinius et al.}
\titlerunning{Nearby nonaccreting black hole}
\abstract{
Several dozen optical echelle spectra demonstrate that \object{HR\,6819} is a hierarchical triple.  A classical Be star is in a wide orbit with an unconstrained period around an inner 40 d binary consisting of a B3\,III star and an unseen companion in a circular orbit.  The radial-velocity semi-amplitude of 61.3\,km/s of the inner star and its \textcolor{black}{minimum} (probable) mass of 5.0\,M$_\odot$ ($6.3 \pm 0.7$\,M$_\odot$) imply a mass of the unseen object of $\geq 4.2 $\,M$_\odot$ ($\geq 5.0 \pm 0.4$\,M$_\odot$), that is, a black hole (BH).  The spectroscopic time series is stunningly similar to observations of LB-1.  A similar triple-star architecture of LB-1 would reduce the mass of the BH in LB-1 from $\sim70$\,M$_\odot$ to a level more typical of Galactic stellar remnant BHs.  The BH in HR\,6819 probably is the closest known BH to the Sun, and together with LB-1, suggests a population of quiet BHs.  Its embedment in a hierarchical triple structure may be of interest for models of merging double BHs or BH + neutron star binaries.  Other triple stars with an outer Be star but without BH are identified; through stripping, such systems may become a source of single Be stars.  
}
%
\keywords{Binaries  --  Stars: black holes  --  Stars: 
  individual: HR\,6819, ALS\,8775 (LB-1)}
\maketitle
%

\section{Introduction}
\label{sect_intro}
The largest discrepancy between Galactic population-synthesis models and
observations probably concerns the number of black holes (BHs).  Models
predict 10$^8$ to 10$^9$ stellar mass BHs \citep{2002MNRAS.334..553A,
  2019arXiv190808775O}, but merely a few $10^2$ X-ray binaries are known, most
of which only harbor neutron stars (NS).  This comparison is severely biased
because it only includes actively accreting BHs.  Nonaccreting BHs are much
more difficult to find \citep[for a recent report of a detection, see][]{2019Sci...366..637T}.  
Compared to plain \textcolor{black}{single-lined spectroscopic binaries (SB1)}, the presence of a second luminous star as a distant third body in a hierarchical triple can provide further constraints on the parameters of nonaccreting BHs \citep{2019arXiv190507100H}.  Massive hierarchical triple systems recently attracted interest because the outer object might accelerate the merger of inner double-BH or BH + NS binaries through Lidov-Kozai oscillations, as detected through gravitational waves \citep[][and references therein]{2019ApJ...882L..24A}.  

\begin{figure}[t]
   \centering
   \includegraphics[width=0.5\textwidth]{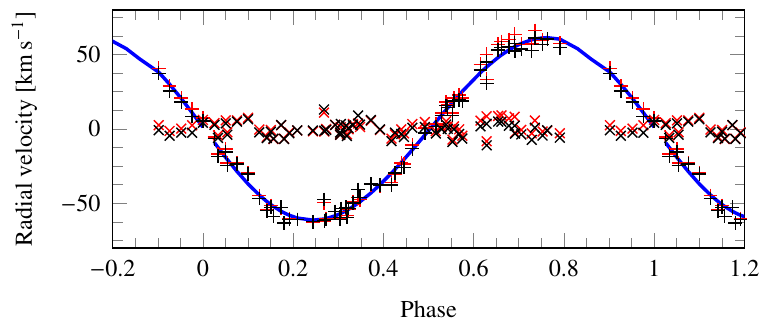}
   \caption{\label{fig_HR6819_phased} The RV curve (black $\text{pluses}$:  \ion{He}{i}$\lambda$4026, red $\text{pluses}$:  \ion{Mg}{ii}$\lambda$4481) of the narrow-line star \textcolor{black}{(see\ Appendix\ \ref{app_RV} and Fig.\,\ref{appfig_HR6819_trails})} in the inner binary, the orbital fit (blue line; Table~\ref{tab_comparison}), and the residuals ($\text{crosses}$).}
\end{figure}

\begin{figure*}[t]
   \centering
   \includegraphics[width=\textwidth]{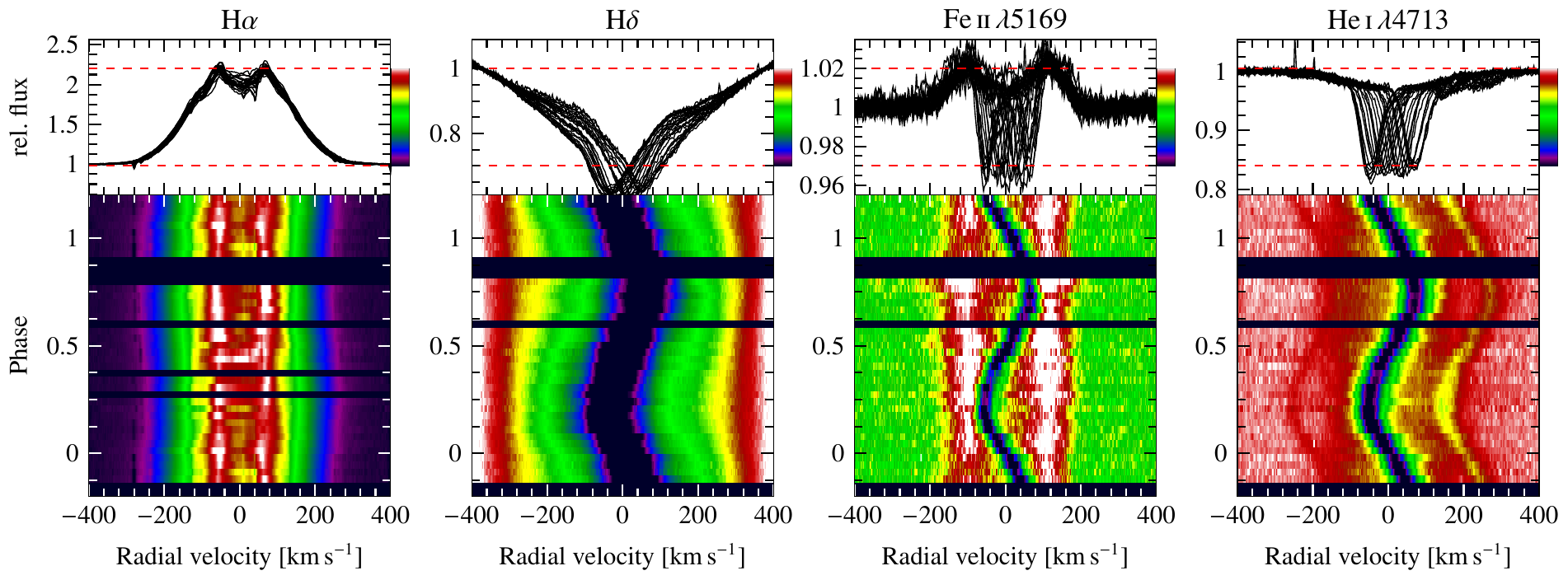}
   \caption{\label{fig_HR6819_trails}
   Dynamical spectra of selected circumstellar and photospheric lines (as labeled) of the HR\,6819 system, phased with the orbital period (see Table~\ref{tab_comparison}) of the inner binary. The individual line profiles are overplotted at the top of each panel; the dashed horizontal lines identify the range of the color-coding in the dynamical spectra.  The \mbox{H$\alpha$}\  panel was constructed with data set B alone (see Table~\ref{tab_specobs}) to minimize the effect of the long-term disk variability of the Be star on the appearance of the emission. The basically invariant RV of the Be star is clearly visible in the peaks of its circumstellar emission lines (e.g., \ion{Fe}{ii}) as well as in the broad, weak absorption bases of predominantly photospheric lines such as \ion{He}{i}.   The cuts in the H$\delta$ panels were chosen to highlight that at the 5\% level (one tick mark in the upper panel) no feature moves in antiphase with respect to the line core. See Fig.~\ref{appfig_HR6819_trails} for additional spectral lines.
   }
\end{figure*}

\begin{table*}[t]
\begin{minipage}{\textwidth}
\begin{center}
\caption[]{\label{tab_specobs} Spectroscopic datasets.  They are available from the LSW Heidelberg \footnote{Set A: \url{https://www.lsw.uni-heidelberg.de/projects/instrumentation/Feros/ferosDB/search.html}} and the ESO Science Archive Facility\footnote{Set B: \url{http://archive.eso.org/cms.html}}, respectively. 
}
\begin{tabular}{ccc@{\,--\,}cllcccc}
\hline\noalign{\smallskip}
\hline
\rule{0ex}{2.5ex} Data  & Observing & \multicolumn{2}{c}{JD}  &   &   &
 \multicolumn{1}{c}{Number\ of} & \multicolumn{1}{c}{Resolving} & \multicolumn{1}{c}{Spectral}   &   \\
set                     & season   & \multicolumn{2}{c}{--2,400,000}   & \raisebox{1.5ex}[1.5ex]{Telescope}
& \raisebox{1.5ex}[1.5ex]{Instrument} &
\multicolumn{1}{c}{spectra}   &\multicolumn{1}{c}{power}  & \multicolumn{1}{c}{range [\AA]}& \raisebox{1.5ex}[1.5ex]{Ref.} \\
\hline \noalign{\smallskip}
A & 1999 & 51\,373 & 51\,394 & ESO 1.52m & \mbox{\sc Feros} &  12 & 48\,000 & 3700-9000 & 1, 2 \\
B & 2004 & 53\,138 & 53\,273 & ESO/MPG 2.2m & \mbox{\sc Feros} &  51 & 48\,000 & 3700-9000 &  2 \\
\hline\noalign{\smallskip}
\end{tabular}
\end{center}
\end{minipage}
References: 1 - \citet{MPhD}, 2 - this work
\end{table*}

The system \object{ALS\,8775} (\object{LB-1} in the following) was recently reported to host an apparently nonaccreting BH with the unusually high mass of $\sim70$\,M$_{\odot}$ \citep{2019Natur.575..618L}. This is not easily reconciled with binary evolution and Galactic population-synthesis models \citep{2019arXiv191203599E}. While the validity of the mass determination involving emission lines has been questioned \citep[e.g., ][]{2019arXiv191204092A, 2020MNRAS.493L..22E}, a BH of more regular mass can still be inferred from the binary mass function derived from the luminous component.  However, none of the rebuttals so far have offered a convincing explanation for the lack of motion of the broad H$\alpha$ emission line, which in fact shows a profile that is very typical for Be stars.

\object{HR\,6819} (also known as HIP\,89605, HD\,167128, or QV\,Tel) appears as a bright (V=5.3\,mag) early-type Be star.  Classical Be stars \citep{2013A&ARv..21...69R} are  extremely rapid rotators, and their emission lines originate from a rotationally supported equatorial disk.
\citet{1981A&AS...43..427D} reported narrow absorption lines of \ion{Ca}{ii} and \ion{Si}{ii}  as well as in \ion{He}{i}$\lambda$4471{\AA, which}  are uncommon in Be stars, and \citet{1982ApJS...50...55S} noted a similarity of the absorption spectrum to that of a normal sharp-lined B3 giant. In spectra from 1999, \citet{MPhD} found that these narrow lines actually \emph{\textup{are}} the signature of a second star with variable radial velocity (RV). 
Because of the small number of observations, she was only able to propose an orbital period of a few dozen days. 
\textcolor{black}{In 2009, Hadrava, \v{S}tefl, and Rivinius 
\textcolor{black}{disentangled} the spectra of the two stars from a larger dataset,  using the method of \citet{1995A&AS..114..393H, 2004PAICz..92...15H}, and discovered that HR\,6819 contains a third, unseen object that may be a BH. The results of this analysis will be presented in a forthcoming paper (in the following referenced as Hadrava et al., in prep.).}

This Letter follows a more conventional and alternative approach. Sect.\,\ref{sect_obsanalysis} derives the orbital parameters of the inner binary and estimates the space trajectory of HR\,6819 as a whole relative to the Solar System.  Section\,\ref{sect_discussion} argues that the unseen component is a BH, compares HR\,6819 to other hierarchical triple systems with an outer Be star, identifies the similarity of the HR\,6819 and LB-1 systems in all architectural details, and briefly sketches the history of HR\,6819 relative to Sco OB2.  
 
\section{Observations and their analysis}
\label{sect_obsanalysis}
\subsection{Time-resolved spectroscopy and photometry}
\label{sect_observations}
The findings by \citet{MPhD} from dataset A in Table~\ref{tab_specobs} motivated further observations of HR\,6819 for several months in 2004 (dataset B in Table~\ref{tab_specobs}). 
The instrument used was the echelle-format {\it Fibre Extended Range Optical Spectrograph} \citep[{\sc Feros},][]{1999Msngr..95...8K}.
The spectra were reduced with the standard {\sc Feros}  pipeline\footnote{\url{https://www.eso.org/sci/facilities/lasilla/instruments/feros/tools/DRS.html}}.  

HR\,6819 is photometrically variable with a full range of nearly 0.1\,mag.  Time series from 
three different space photometers are described and discussed in Appendix~\ref{app_photometry}.  No regularly repeating variability was detected.  

\subsection{Inner orbit}
Figure\,\ref{fig_HR6819_phased} and Appendix\,\ref{RVdata} present the  RVs of the sharp-lined star for which the {\sc Fotel} code of \citet{2004PAICz..92....1H} yielded the orbital parameters of the inner binary compiled in Table\,\ref{tab_comparison}.  
\textcolor{black}{Because dataset B spans several cycles, the orbital period is easily constrained to be about 40.3\,d based on datasets A and B. {\sc Fotel} gives a period of $40.333\pm0.004$\,d. A further and more detailed analysis that also includes historical data and evaluates the possibility of secular changes, will be provided by Hadrava et al., in prep.
}
Because of the low eccentricity, the \textcolor{black}{argument of the periastron, $\omega$,} is poorly constrained ($\pm20^\circ$), and its errors propagate into those of \textcolor{black}{the velocity amplitude, K$_1$}.  Therefore $\omega$ was fixed, whereupon the errors on K$_1$ dropped to the 1\% level.  In search for a possible orbital motion between 1999 and 2004 of the inner binary relative to the outer Be star, a third iteration also required common results from datasets A$+$B, but permitted the \textcolor{black}{systemic velocity, $\gamma$,} of the inner binary to be different for A and B.  The difference $\gamma_{\rm 1999}-\gamma_{\rm 2004} \approx +2$\,km\,s$^{-1}$ is not significant, in agreement with the positional invariance of the emission lines from the outer Be star (Figs.\,\ref{fig_HR6819_trails} and \ref{appfig_HR6819_trails}; Hadrava et al., in prep.).  Because the outer orbit is unknown, \textcolor{black}{but may have a period of some decades and any orientation, an undetectably small orbital acceleration over a few years is very well possible, and} the RV of the complete triple system is preliminarily adopted as the combined solution of sets A$+$B, $\gamma = +9.4 \pm0.5$\,km\,s$^{-1}$.  

\subsection{Spatial position and trajectory of the triple system}
\label{sect_trajectory}
Neither the Gaia  \citep[DR2][$\pi=2.915$\,mas]{2018yCat.1345....0G} nor the Hipparcos \citep[][$\pi=4.29$\,mas]{2007A&A...474..653V} parallax solutions for HR\,6819 consider binarity; they are therefore highly uncertain. 
The Gaia measurements may suffer from the brightness of HR\,6819.  However, at a distance of about 310\,pc (see below), the angular diameter of the circular orbit of the B3\,III star of at least 0.44\,au (for $\sin i=1$) 
is 1.4\,mas or more. 
This matches the discrepancy as well as the astrometric excess noise of 0.731\,mas in the Gaia solution. The observed flux is approximately compatible with both parallaxes.  Depending on other assumptions made, a preference for a distance of about $310\pm60$\,pc emerges from flux fitting (Appendix\,\ref{app_distance}). 

The proper motion (in mas\,yr$^{-1}$: $-3.667$ and $-11.120$ from Gaia and $-4.10$ and $-13.30$ from Hipparcos in $\alpha$ and $\delta$, respectively) is much larger than the parallax and thus is less disturbed by any orbital pattern. At a distance of 310\,pc, 1\,mas\,yr$^{-1}$ translates into 1.5\,km\,s$^{-1}$. When we average the Gaia and Hipparcos velocities as a simple estimate, the total lateral velocity is 12.8\,\mbox{km\,s$^{-1}$}, or 13.1\,pc\,Myr$^{-1}$ , and the radial velocity of 9.4\,\mbox{km\,s$^{-1}$}\ corresponds to 9.6\,pc\,Myr$^{-1}$. Without taking the 230\,Myr solar orbit in the Galaxy into account, that is, when we assume that the corotating frame is inertial, the closest approach of HR\,6819 to the Sun is estimated for 11\,Myr ago, at a distance of about 260\,pc.  A more precise trajectory and origin can only be derived when the outer orbit is known and the astrometric parameters have been computed taking the multiplicity into account.

\begin{table}
\caption{
Parameters of the inner binary in HR\,6819 compared to those of the inner binary in LB-1 \citep{2019Natur.575..618L}.  \textcolor{black}{The minimum BH mass for HR\,6819 was derived with $f_{\rm M}-2\sigma = 0.90$ instead of $f_{\rm M}$ itself to obtain a true lower limit.} 
}
\label{tab_comparison}      
\centering                           
\begin{tabular}{l c c}          
\hline\hline                        
Parameter \hfill [unit] & HR\,6819 & LB-1 \\    
\hline                            
$P$ \hfill [d]                  & $40.333\pm0.004$ & 78.9 \\
$K_1$ \hfill [km\,s$^{-1}$]       & $61.3\pm0.6$   & $52.8$ \\
$e$                             & $0.03\pm0.01$  & $0.03$ \\
$T_{\rm conj}$     \hfill [MJD]        & $53177.44\pm0.07$ & N/A \\
$\omega$ (fixed) \hfill [$^{\circ}$]    & 89  & N/A\\
$\gamma$ \hfill [km\,s$^{-1}$]  &  $9.4\pm0.5$ & $28.7$\\
Mass function $f_{\rm M}$ \hfill [M$_\odot$]            & $0.96\pm0.03$   & 1.20\\
Minimum B star mass \hfill [M$_\odot$]        &  5.0    & 8.2  \\
Minimum BH mass \hfill [M$_\odot$]   & $> 4.2$  & $>6.3$ \\
\hline                                      
\end{tabular}
\end{table}

\section{Discussion}
\label{sect_discussion}
\subsection{Nature of the unseen inner component}
\label{sect_innerorbit}
With the above orbital parameters, the mass function of the unseen component
in the inner system becomes $0.96 \pm 0.03$\,\mbox{${\rm M}_{\odot}$}
\textcolor{black}{(see Appendix~\ref{app_RV} for details)}. Evaluating a lower limit on the mass of the unseen component requires a lower limit on the mass of the B3\,III star.  
Appendices\,\ref{app_minmass} and \ref{app_bestfit} demonstrate that B3\,III is sound as a spectral classification and agrees with the usage of that spectral type by \citet{2010AN....331..349H}.  The database compiled by these authors contains 56 mass determinations for B3\,III stars, none of which has a median value $<$5.0\,\mbox{${\rm M}_{\odot}$}. This lower limit is also in agreement with evolutionary tracks (Appendix\,\ref{app_minmass}).  When we assume a lower limit of 5 \mbox{${\rm M}_{\odot}$}\ for the B3\,III star and the 3$\sigma$ lower limit (0.87\,M$_\odot$) on the mass function, a hard lower limit on the mass of the unseen object is 4.2\,M$_\odot$.

Next, we need to verify whether a normal star with a mass at the lower limit can hide in the spectra of HR\,6819.  According to \citet{2010AN....331..349H}, the spectral type of stars with a \emph{\textup{typical}} mass of 4.2\,\mbox{${\rm M}_{\odot}$}\ is about B7.  For the lowest possible luminosity class of V, and following \citet{2006MNRAS.371..185W}, the absolute visual magnitudes $M_V$ of a B3\,III and a B7\,V star differ by 1.7\,mag, or a factor of almost 5 in flux at 550\,nm. Because the B and Be star both contribute about equally to the total $V$-band flux (Appendix~\ref{app_bestfit}; Hadrava et al., in prep.), $\text{about }$10\% of the continuum would come from the inner companion if it were luminous. For this relative brightness, modeling with synthetic spectra from \citet{2018A&A...609A.108S} shows that a B7\,V star would appear with a spectral signature of 0.6\% in \ion{Mg}{ii}$\lambda$4481 and 6\% in the Balmer lines even
at near-critical rotation with $v \sin i > 200$\,km\,s$^{-1}$ . That is, these features would not have gone unnoticed in the dynamical spectra that were constructed from more than 50 {\sc Feros} spectra with an average signal-to-noise ratio $S/N > 280$ at 450\,nm, and the unseen component is not a normal star.  \cite{2018ApJ...853..156W} searched UV spectra for an sdO companion.  However, the $S/N$ of the calculated cross-correlation function was too low for a detection.  \textcolor{black}{Because the mass difference between the B star and the unseen object increases with the mass of the B star, the maximum mass of the unseen object also needs to be evaluated in principle because it might lead to a larger magnitude difference and hence lower detectability of a luminous companion. For HR\,6819, however, this is not the case (see Appendix\ \ref{app_maxmassdiff}).}

HR\,6819 is not positionally coincident with any known pulsar\footnote{The ATNF catalog \citep{2005AJ....129.1993M} is maintained at \url{https://www.atnf.csiro.au/research/pulsar/psrcat/ }}.  
The lower limit of 4.2\,\mbox{${\rm M}_{\odot}$}\ on the mass of the unseen
object is substantially above the empirical mass limit of $\sim$2.6\,M$_\odot$
for NSs \citep[][\textcolor{black}{moreover, a companion mass close to that
    limit would imply an unbelievably
low mass for the B star, see Appendix\ \ref{app_RV}}]{2018MNRAS.478.1377A} and also falls into the possible gap between observed masses of NSs and BHs \citep[see ][]{2011ApJ...741..103F}.
Therefore, the unseen object must be a BH.

Black holes are traditionally detected in X-rays that result from the accretion of matter from a companion star. 
\textcolor{black}{By contrast, in the only recent X-ray observation of HR\,6819, the {\it ROSAT} all-sky survey, the source is not detected \citep[$L_X \leq 3.5 \times 10^{30}$\,erg/s in the 0.1-2.4 keV band, assuming a distance of 460 pc;][]{1996A&AS..118..481B}, and the near-circularity of the orbit does not suggest major variability in the form of undetected outbursts, as is confirmed in optical light (Appendix\,\ref{app_photometry}).}  Accordingly, the inner binary is not interacting.  At the temperature and luminosity of a B3\,III star, radiative winds are intrinsically very weak \citep{2014A&A...564A..70K} and are ineffective over a separation of $>0.22$\,au (the lower limit from the circular inner B3\,III-star orbit alone).  Therefore, the lack of interaction is expected.  This suggests that for all practical purposes, the BH is not accreting. 

If, instead of the minimum
mass of 5.0\,M$_\odot$ and the 3$\sigma$ lower limit on the mass function, we use the more typical mass of a B3\,III star of $6.3 \pm 0.7$\,M$_\odot$ (Appendix\,\ref{app_bestfit}) and the nominal value of the mass function, $0.96 \pm 0.03$\,M$_\odot$,  the BH mass increases from 4.2\,M$_\odot$ to $5.0 \pm0.4$\,M$_\odot$. For orbital inclinations $<90^\circ$ (the angle is unconstrained), both values would be higher.  

\subsection{Similar systems}

As already noted by \citet{2019Natur.575..618L},
the shape of the line emission in LB-1 closely matches that of Be stars, although these authors only speak of 'wine-bottle profiles' \citep[which are typical for Be stars, see][]{1996AAS..116..309H}.   The undetected RV amplitude of the emission line \citep[][]{2020MNRAS.493L..22E} is naturally explained by a Be star at a large distance from the other luminous star, as in HR\,6819. Table\,\ref{tab_comparison} compares the properties of the inner binaries in HR\,6819 and LB-1. A forthcoming paper (Heida et al., in prep.) will \textcolor{black}{disentangle}
the spectra of the two luminous stars in LB-1, which leads to a drastic reduction of the BH mass \citep[see][]{2019arXiv191204092A, 2020MNRAS.493L..22E}.

In principle, both HR\,6819 and LB-1 might be chance superpositions of a binary and a Be star, although this would stretch simple probability considerations.  In both systems, the absolute and apparent spectroscopic luminosities of the visible components are similar (HR\,6819:  Appendix\,\ref{app_distance}, Hadrava et al., in prep.; LB-1:  Heida et al., in prep.), and the similar distances support the assumption that both are physical triples.  

The great similarity of HR\,6819 and LB-1 invites the question whether this is enough to identify possible progenitor systems.  The simplest assumption is that the hierarchy was not altered by the formation of the BH, whereas the initial inner binary was tighter and had a total mass of well over $\text{}$10\,M$_\odot$ \citep{2016ApJ...821...38S}.  One such candidate system may be the detached eclipsing 2.7 d binary \object{CW\,Cep} with masses of 13 and 12\,M$_\odot$ and a near-circular orbit \citep{2019A&A...628A..25J}.  Whether the two stars can avoid merging strongly depends on the mass loss.  \citet[][their Fig.\,2]{2019A&A...628A..25J} reported broad H$\alpha$ emission that does not participate in the orbital motions, but has the typical profile of classical Be stars.  This means that in CW\,Cep, an outer Be star may also be orbiting an inner binary.  The H$\alpha$ profile indicates an intermediate inclination of the Be star \citep[see][]{2000A&A...359.1075H}, while the orbital inclination of the inner binary is around zero because the system is eclipsing.  

At first sight, it seems perplexing that all three systems host an outer Be star.  Because HR\,6819 and CW\,Cep were identified because their known line emission is similar to that in LB-1, this is probably a selection bias.   However, \object{66\,Oph} is a fourth hierarchical triple with an outer Be star \citep[but without a BH,][]{2004IAUS..215..166S}.  Such a channel for the formation of Be stars \citep[for a description of other channels, see][]{2020A&A...633A..40L} may be due to angular momentum distribution during fragmentation in massive-star formation \citep{1978ApJ...224..488B, 2006MNRAS.373.1563K}.  If these outer Be stars become unbound, they may have above-average space velocities, but probably not at the top of the range observed in a fair fraction of Be stars \citep{2001ApJ...555..364B}.  Another conspicuous commonality is the vanishing eccentricity of the inner binaries, which is not typical of hierarchical triple systems \citep{2019AJ....158..167T}.  

\subsection{Origin and history of HR\,6819}
The Solar System is located within the so-called Local Bubble, which measures about 100\,pc across and is characterized by a lower-than-average density, but high temperatures of its gas content \citep{2017hsn..book.2287S}.  This structure is attributed to supernovae (SNe) that exploded several million years ago \citep[as summarized by][]{2017hsn..book.2253F} when Sco OB2, the most nearby stellar \textcolor{black}{OB} association, passed in the solar vicinity \citep{2006MNRAS.373..993F}.  The analysis in Appendix~\ref{app_Bestar} shows that HR\,6819 is probably older than Sco OB2, and according to Sect.\,\ref{sect_trajectory}, its trajectory does not cross that of Sco OB2 \citep{1999AJ....117..354D}, which therefore is not the parent population of HR\,6819.  Crude preliminary estimates of the distance of HR\,6819 at the time of the SN explosion range from 280\,pc if the explosion occurred 15\,Myr ago to about 800\,pc for an explosion $\sim$\,65-70\,Myr ago (see Appendix~\ref{app_Bestar} for a discussion of the age) and place this event well outside the Local Bubble.  Depending on the initial mass, it is also  possible that the BH formed without explosion by direct collapse \citep{2016ApJ...821...38S}.

\section{Conclusions}
\label{concl}
HR\,6819 is a hierarchical triple star with a nonaccreting BH in the inner binary.  The mass estimate does not depend on difficult RV measurements of emission lines as tracers of the orbital motion of the BH \citep[see][]{2014Natur.505..378C, 2020MNRAS.493L..22E}.  Furthermore, the B3\,III star is a fairly normal star so that the combination of the mass derived from spectral type and \textcolor{black}{spectral energy distribution} with the mass function leads to a solid lower mass limit on the BH of 4.2\,M$_\odot$ (for $\sin i = 1$).   
\textcolor{black}{The \object{CW\,Cep} system} may be a progenitor of systems with an architecture similar as in \object{HR\,6819} and \object{LB-1} if the 2.7 d inner binary can avoid merger through mass loss.   Through stripping of the outer Be star, systems such as HR\,6819, LB-1, CW\,Cep, and 66\,Oph may become a source of single Be stars.  

The detection was facilitated by the incompatibility of the spectral sequences with a simple binary.  This can be a criterion to find other nonaccreting BHs, and the case of LB-1 illustrates the scope for such searches.  
If $\text{about }$20\% of all early-type stars are triples \citep[see\ ][]{2017ApJS..230...15M} and 0.01\% of them have a system architecture similar to that of HR\,6819 and LB-1, the discrepancy between expected and observed numbers of BHs in the Galaxy would shrink by several orders of magnitude but would still be very large. \citep[In the magnitude-limited Bright Star Catalog,][HR\,6819 corresponds to 0.01\% of all early- and late-type stars together.]{1991bsc..book.....H}  

The existence of an entire population of quiet BHs is also suggested by the relative proximity of HR\,6819.  The lowest distance derived to date to any Galactic BH apparently is $<400$\,pc, which is the distance to the accreting  \object{V616\,Mon} \citep[AO\,0620-00;][]{2009NewA...14..674F}.  The Gaia parallax of $0.64\pm0.16$\,mas is closer to other distance estimates of $\sim1$\,kpc \citep{2009NewA...14..674F}, but would still make V616\,Mon the nearest previously known BH to the Sun.  At $310\pm60$\,pc, the distance to HR\,6819 is far smaller than 1\,kpc.  

Triple systems have been invoked \citep{2018ApJ...863....7R} as the progenitors of double BHs or BH + NS systems that merged owing to Lidov-Kozai oscillations triggered by a distant third object and were detected as gravitational-wave events \citep{2019ApJ...882L..24A}.  It would therefore be interesting to compare dynamical models to  HR\,6819 or LB-1.  However, the two luminous components of HR\,6819 have masses of only $\sim$\,6\,M$_\odot$ and are very far apart, therefore the current BH will remain the only BH in HR\,6819, and neither will an NS form.  It appears noteworthy that the pure binary models of \citet{2019arXiv191209826L} predict a steep decrease in the fraction of BH companions toward the masses of the inner B stars in HR\,6819 and LB-1.  It may be useful to extend the models to triple systems.  

HR\,6819 does not share the proper motion of the \textcolor{black}{young stellar associaton} Sco OB2 and is probably also older than it.  Any SN explosion leading to the BH formation most probably occurred outside the Local Bubble.  

A subsequent paper (Hadrava et al., in prep.) will present the results of \textcolor{black}{disentangling}
the spectra.  The results are expected to achieve a comprehensive and largely independent quantitative confirmation of the properties of HR\,6819 that we presented here.

\begin{acknowledgements}

This work was started when PH was a visitor at ESO, Santiago. MH is supported by an ESO fellowship. RKl acknowledges support by the National Science Foundation under Grant No.~AST-1908026. This research has made use of NASA’s Astrophysics Data System \citep[ADS,][]{2000A&AS..143...41K} and the SIMBAD database \citep{2000A&AS..143....9W} operated at CDS, Strasbourg, France. \textcolor{black}{We are grateful for the referee's suggestion, which in particular led to an improved discussion in the appendices.}

\end{acknowledgements}

\bibliography{hr6819.bib} 


\appendix

\section{Radial velocities of the narrow-line inner B3\,III star 
\textcolor{black}{and the mass function}}\label{app_RV}
Table\,\ref{RVdata} lists the RVs measured by fitting single Gaussians to the narrow cores of the \ion{He}{i}$\lambda$4026 and \ion{Mg}{ii}$\lambda$4481 lines.  These cores move across the broad absorption profiles of the Be star, which are stationary at about the mean velocity of the B3\,III star.  Because this background flux increases toward both positive and negative velocities, the measured RV swing may be slightly reduced, which would tend to decrease the inferred minimum mass of the BH.

The mass function \textcolor{black}{\citep[e.g.,][]{1973bmss.book.....B}}
\begin{equation}\label{eq_massfunction} 
f_{\rm M} \equiv \frac{PK_1^3}{2\pi G} (1-e^2)^{3/2} = \frac{M^3_{\rm BH}}{(M_{\rm B} + M_{\rm BH})^2 } \sin^3 i
\end{equation}
can be solved iteratively for $M_{\rm BH}$ as
\begin{equation}\label{limitmass}
\lim_{n\to\infty} M_{{\rm BH},n+1} = (M_{\rm B} + M_{{\rm BH},n})^{2/3} f_{\rm M}^{1/3}  \sin^{-1}i
.\end{equation}
Considering the given values \textcolor{black}{(see Table~\ref{tab_comparison})}, this means that $M_{\rm BH}$ decreases with $M_{\rm B}$, but more slowly. For the following discussion, $\sin i=1$ is assumed to provide a limit. Evaluation of Eq.\,\ref{limitmass} for a grid of hypothetical values of $M_{\rm B}=[3,5,7,9]$\,\mbox{${\rm M}_{\odot}$}\  (which fully brackets the range for the B star, considering its spectrum, see below), returns minimum values of $M_{\rm BH} = [3.4,4.4,5.2,6.0]$\,\mbox{${\rm M}_{\odot}$}, respectively. In other words, for any plausible B star mass, the \emph{\textup{minimum}} mass ratio $M_{\rm BH}/M_{\rm B}$ is always close to unity, and for a B-star mass below about 3.8\,\mbox{${\rm M}_{\odot}$}, even rises to above one in favor of the unseen companion.

\textcolor{black}{The analysis can be turned around by asking  whether the minimum conventional BH mass of 3\,\mbox{${\rm M}_{\odot}$}\ can be reconciled with a realistic mass of the B3\,III star.
For this, Eq.\,\ref{eq_massfunction} can be solved  for $M_{\rm B}$ directly as
\begin{equation}
M_{\rm B} = \sqrt{\frac{M_{\rm BH}^3 \sin^3 i}{f_{\rm M}}} - M_{\rm BH}
.\end{equation}
Substituting 3\,\mbox{${\rm M}_{\odot}$}\ for $M_{\rm BH}$ 
leads to $M_{\rm B}=2.3$\,\mbox{${\rm M}_{\odot}$}\ (for $\sin i=1$, and even smaller for lower inclinations), a value that can be safely excluded because it is well below any conceivable mass of a B3\,III star. As a consequence, if the companion does not produce nuclear energy in its core, and hence is luminous, it must be a BH and cannot be an NS or even a white dwarf. The discussion in Appendices \ref{app_minmass} and \ref{app_maxmassdiff}  therefore mainly focuses on the question of whether a luminous companion could have gone undetected.}

\begin{table*}
\caption{Barycentric radial velocities of the inner B3\,III star.  BJD is Julian Date -- 2,400,000.5}
\label{RVdata}
\centering                           
\begin{tabular}{rrrrrrrrrl}          
\hline\hline
BJD \hspace{5mm} & \multicolumn{2}{c}{Radial velocity [km\,s$^{-1}$]\hspace*{-5mm} }&
BJD \hspace{5mm} & \multicolumn{2}{c}{Radial velocity [km\,s$^{-1}$]\hspace*{-5mm} }&
BJD \hspace{5mm} & \multicolumn{2}{c}{Radial velocity [km\,s$^{-1}$]\hspace*{-5mm} }&\\
\hspace{-2mm} &
\multicolumn{2}{c}{\ion{He}{i}$\lambda$4026 \hspace*{2mm} \ion{Mg}{ii}$\lambda$4481\hspace*{-5mm}} & &
\multicolumn{2}{c}{\ion{He}{i}$\lambda$4026\hspace* {2mm}  \ion{Mg}{ii}$\lambda$4481\hspace*{-5mm}} & &
\multicolumn{2}{c}{\ion{He}{i}$\lambda$4026\hspace*{2mm}  \ion{Mg}{ii}$\lambda$4481\hspace*{-5mm}} \\   
\hline  
51373.172 &\hspace*{3mm} $-40.2$ & $-40.1$ &\hspace*{7mm} 53162.357 &\hspace*{5mm} $ 42.8$ & $ 40.1$ &\hspace*{7mm} 53244.049 &\hspace*{3mm} $\hspace*{2mm} 66.2$ & $ 62.4$ \\
51374.170 & $-46.2$ & $-46.3$ &53183.043 & $-44.8$ & $-45.5$ &53245.008 & $ 69.6$ & $ 64.7$ \\
51375.148 & $-44.1$ & $-43.3$ &53185.046 & $-50.8$ & $-51.5$ &53246.117 & $ 66.2$ & $ 63.3$ \\
51376.220 & $-36.8$ & $-31.4$ &53187.161 & $-53.5$ & $-53.0$ &53247.050 & $ 66.4$ & $ 62.1$ \\
51378.178 & $-28.7$ & $-28.4$ &53188.182 & $-52.9$ & $-52.2$ &53248.020 & $ 68.9$ & $ 66.2$ \\
51380.320 & $-14.1$ & $-16.5$ &53190.199 & $-48.8$ & $-50.0$ &53254.038 & $ 50.2$ & $ 46.8$ \\
51383.168 & $ 12.7$ & $ 11.1$ &53194.186 & $-26.8$ & $-28.4$ &53255.013 & $ 38.1$ & $ 34.9$ \\
51384.238 & $ 25.4$ & $ 23.7$ &53195.058 & $-13.4$ & $-16.5$ &53256.054 & $ 30.1$ & $ 27.4$ \\
51385.255 & $ 32.1$ & $ 28.9$ &53196.046 & $ -1.3$ & $ -3.8$ &53257.033 & $ 22.8$ & $ 17.8$ \\
51390.241 & $ 62.2$ & $ 61.7$ &53197.165 & $ 10.0$ & $  6.6$ &53257.991 & $ 16.7$ & $ 14.8$ \\
51393.123 & $ 69.6$ & $ 69.9$ &53199.065 & $ 27.4$ & $ 25.2$ &53259.030 & $  4.0$ & $  2.9$ \\
51394.254 & $ 66.9$ & $ 63.9$ &53202.140 & $ 52.9$ & $ 49.0$ &53260.035 & $ -4.0$ & $ -6.1$ \\
53138.343 & $ -7.4$ & $ -9.0$ &53204.008 & $ 68.2$ & $ 63.9$ &53261.040 & $-13.4$ & $-14.2$ \\
53139.178 & $-13.4$ & $-15.0$ &53205.192 & $ 72.9$ & $ 66.9$ &53262.029 & $-20.1$ & $-21.0$ \\
53143.303 & $-49.5$ & $-49.3$ &53206.996 & $ 75.6$ & $ 70.6$ &53263.035 & $-35.4$ & $-37.3$ \\
53144.249 & $-53.5$ & $-53.7$ &53226.093 & $-51.5$ & $-51.5$ &53264.066 & $-42.2$ & $-43.3$ \\
53149.258 & $-50.8$ & $-50.0$ &53229.987 & $-46.2$ & $-44.0$ &53264.989 & $-43.5$ & $-43.3$ \\
53149.261 & $-51.5$ & $-51.5$ &53231.066 & $-39.5$ & $-40.3$ &53269.008 & $-50.8$ & $-51.5$ \\
53154.164 & $-21.4$ & $-20.2$ &53238.992 & $ 20.1$ & $ 18.5$ &53271.014 & $-44.8$ & $-44.0$ \\
53159.300 & $ 20.1$ & $ 18.5$ &53240.099 & $ 30.1$ & $ 28.2$ &53272.040 & $-38.1$ & $-36.6$ \\
53160.182 & $ 30.1$ & $ 28.2$ &53243.012 & $ 59.5$ & $ 54.5$ &53273.005 & $-28.1$ & $-27.7$ \\
\hline      
\end{tabular}
\end{table*} 

\section{Space photometry of HR\,6819} \label{app_photometry}
 
\begin{figure}[t]
   \centering
   \includegraphics[width=0.5\textwidth]{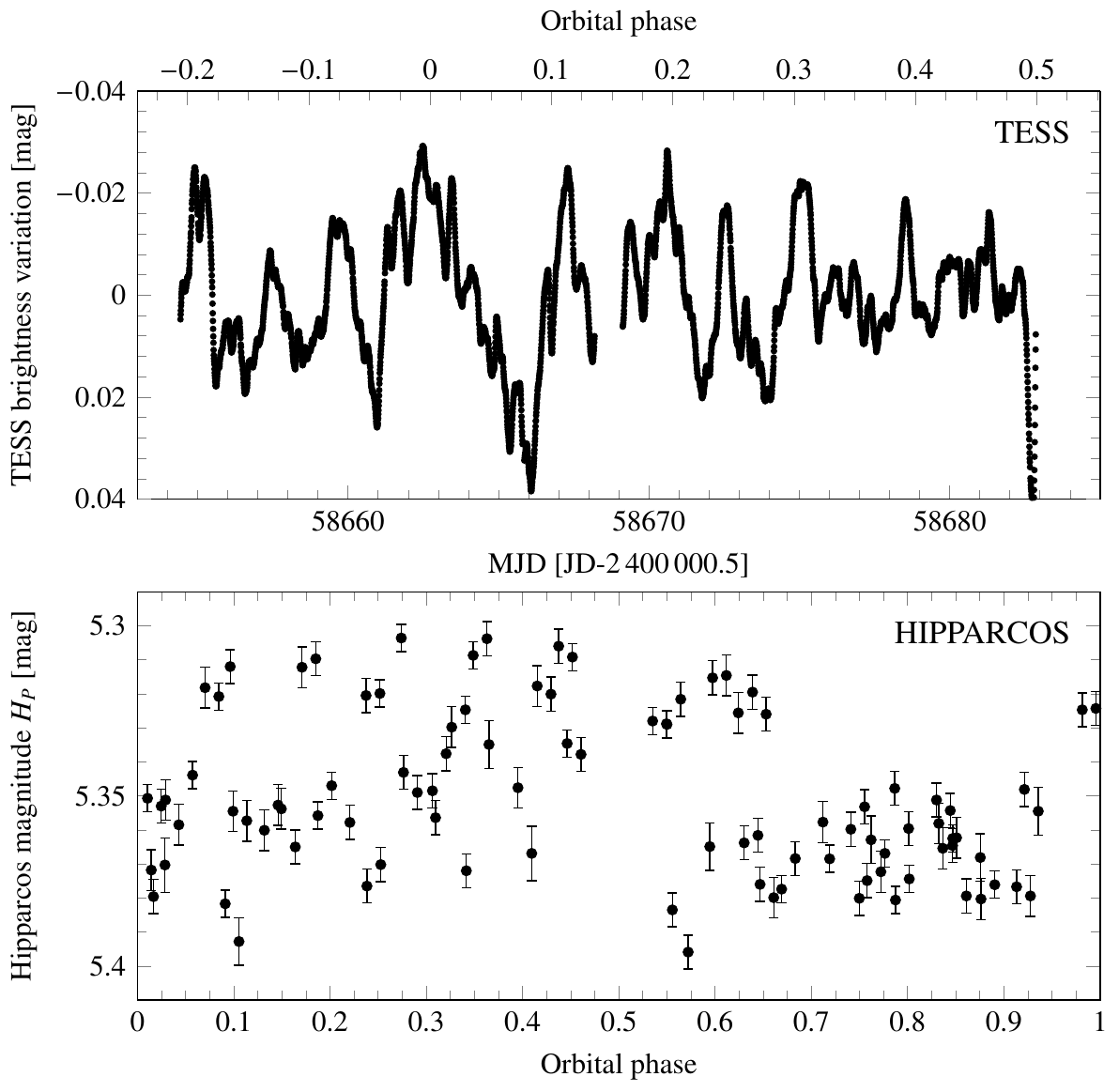}
   \caption{\label{appfig_photometry}Photometric data from TESS (upper panel) and Hipparcos (lower panel, folded with the orbital period of 40.333\,d, see Table~\ref{tab_comparison}). The TESS photometry spans less than one orbital cycle, so that both Julian date and phase are indicated on the axes. }
\end{figure}

\begin{figure}[t]
   \centering
   \includegraphics[width=0.5\textwidth]{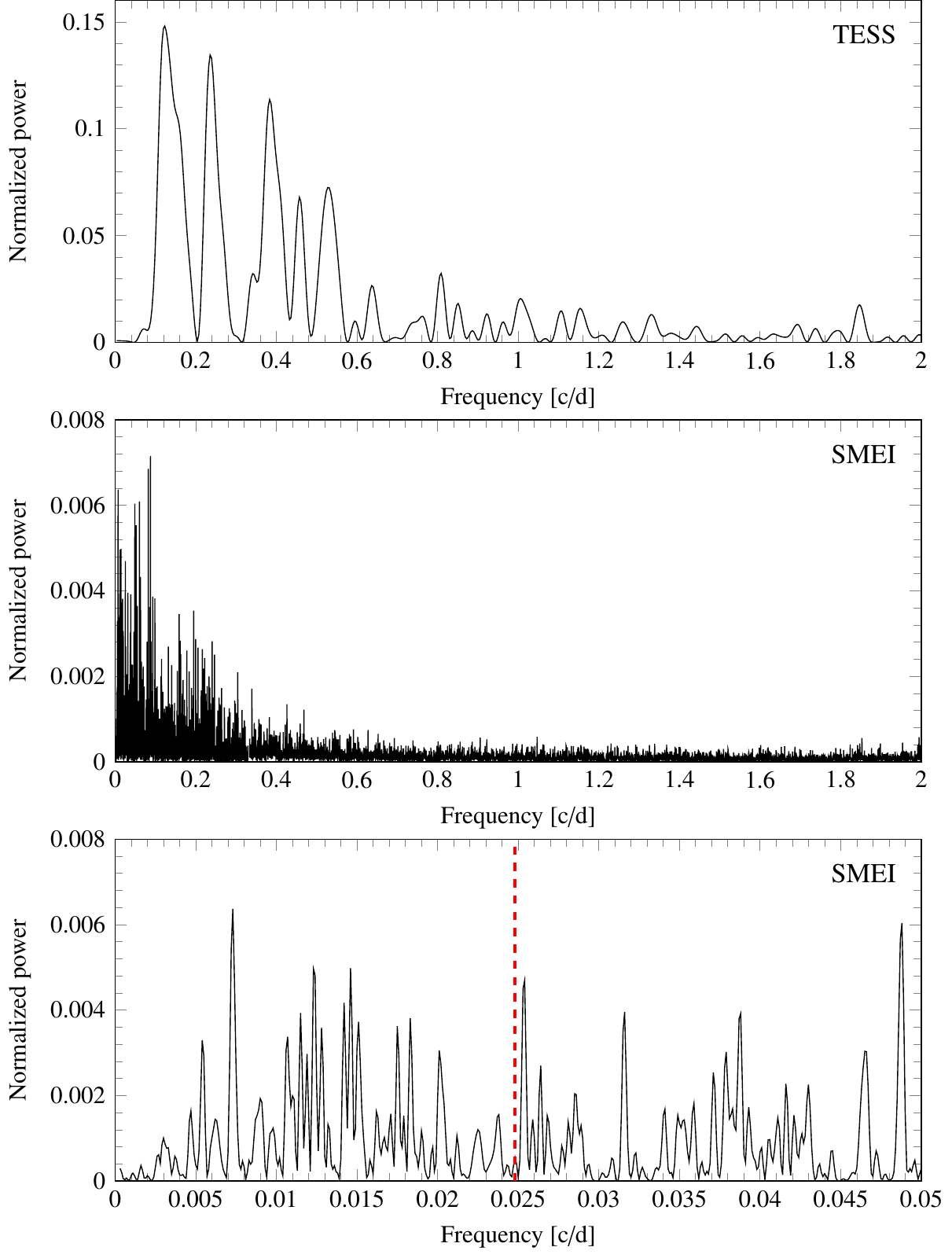}
   \caption{\label{appfig_perig}Periodograms of TESS and SMEI photometry (as labeled). The bottom panel shows the SMEI periodogram around the orbital frequency, marked by a vertical dashed line. }
\end{figure}

HR\,6819 (= QV\,Tel) was observed by the {\it Transiting Exoplanet Survey Satellite} \citep[TESS,][]{2015JATIS...1a4003R} in Sector 13 for $\sim$28.4 days\footnote{Data available from the {\it Mikulski Archive for Space Telescopes} (MAST):
\url{https://dx.doi.org/10.17909/t9-tejq-s571}}. This is too short for a formal period search in the range of the 40 d spectroscopic period (Table~\ref{tab_comparison}).   However, phase-folding the data with this period did not indicate a coherent variability on that timescale. 

On timescales starting at about half a day, the system is variable with a peak-to-peak amplitude of nearly 70\,mmag (Fig.\,\ref{appfig_photometry}, top panel). At B3\,III, the inner B star may be a slowly pulsating B star \citep[SPB star; an example is 18\,Peg (B3\,III):][]{2016A&A...591L...6I} and the outer Be star may also be pulsating \citep{2003A&A...411..229R}. Typical pulsation amplitudes rarely exceed 10\,mmag.   From Be stars, much larger amplitudes are known and can often be traced to matter ejected into the inner regions of the equatorial disk when viewed at small to intermediate angles \citep[intermediate to large inclination angles of the stellar rotation axis;][]{2016A&A...588A..56B, 2018pas8.conf...69B}.  The appearance of the H$\alpha$ emission-line profile (Fig.\,\ref{fig_HR6819_trails}) clearly indicates an orientation of the disk not far from face-on \citep{1994A&A...289..458H, 2013A&ARv..21...69R}.  Because the Be star probably only accounts for little over half of the total optical flux (Appendix~\ref{app_distance}), it would be fairly active, which is not unusual for early-type Be stars \citep[e.g.,][]{2018AJ....155...53L, 2018MNRAS.479.2909B}.  This large-amplitude circumstellar variability probably dominates any pulsational signatures of the two luminous stars (see Fig.~\ref{appfig_perig}, upper panel).  

Hipparcos epoch photometry \citep{1997ESASP1200.....P} does not indicate any sign of the orbital period either and is not suitable to study shorter timescales, even though the variability as such is clearly visible (Fig.\,\ref{appfig_photometry}, bottom panel). Data obtained with the {\it Solar Mass Ejection Imager} \citep[SMEI,][]{2013SSRv..180....1H} paint a similar picture: Although the amplitude observed by TESS and Hipparcos is within the reach of SMEI \citep{2007SPIE.6689E..0CH}, no signal clearly linked to the orbit emerges from the SMEI observations either (see Fig.~\ref{appfig_perig}, lower two panels).

\section{Stellar parameters adopted for the inner B star and the implied distance to the system} 
\label{app_distance}

\subsection{Minimum masses \textcolor{black}{in the inner system}}
\label{app_minmass}

\begin{figure}[t]
   \centering
   \includegraphics[width=0.5\textwidth]{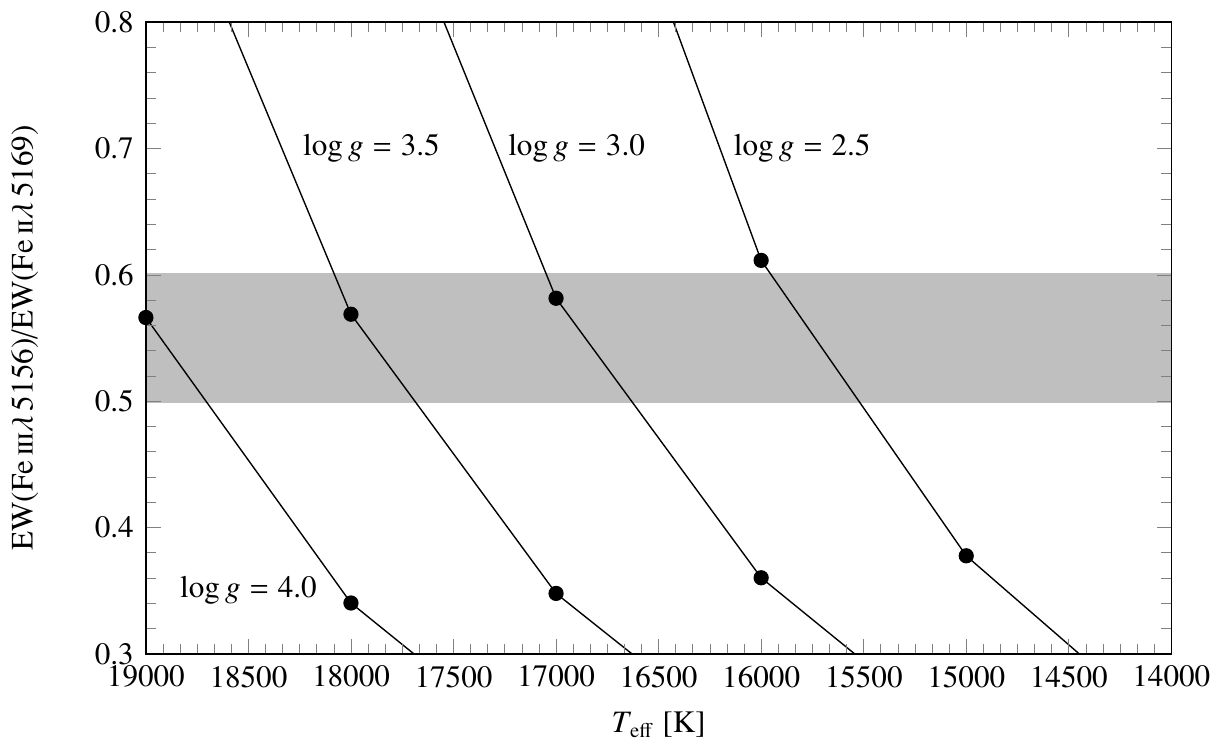}

   \caption{\label{appfig_ionbalance} \ion{Fe}{iii}$\lambda$\,5156 to \ion{Fe}{ii}$\lambda$\,5169 equivalent-width ratio vs.\ effective temperature ($T_{\rm eff}$). Measurements in theoretical spectra \textcolor{black}{\citep[the nonrotating input grid used by][]{2018A&A...609A.108S}} are shown as lines for various values of $\log g$ (as labeled).  The range observed in the inner (narrow line) B star of HR\,6819 is shaded in gray.  This diagnostic diagram suggests that $T_{\rm eff}$ lies between 16 and 18\,kK.  A more precise value is difficult to give owing to blending with the \ion{Fe}{ii}$\lambda$\,5169 emission line of the Be star (see Fig.~\ref{fig_HR6819_trails}). }
\end{figure}

\begin{figure*}[t]
   \centering
   \includegraphics[width=\textwidth]{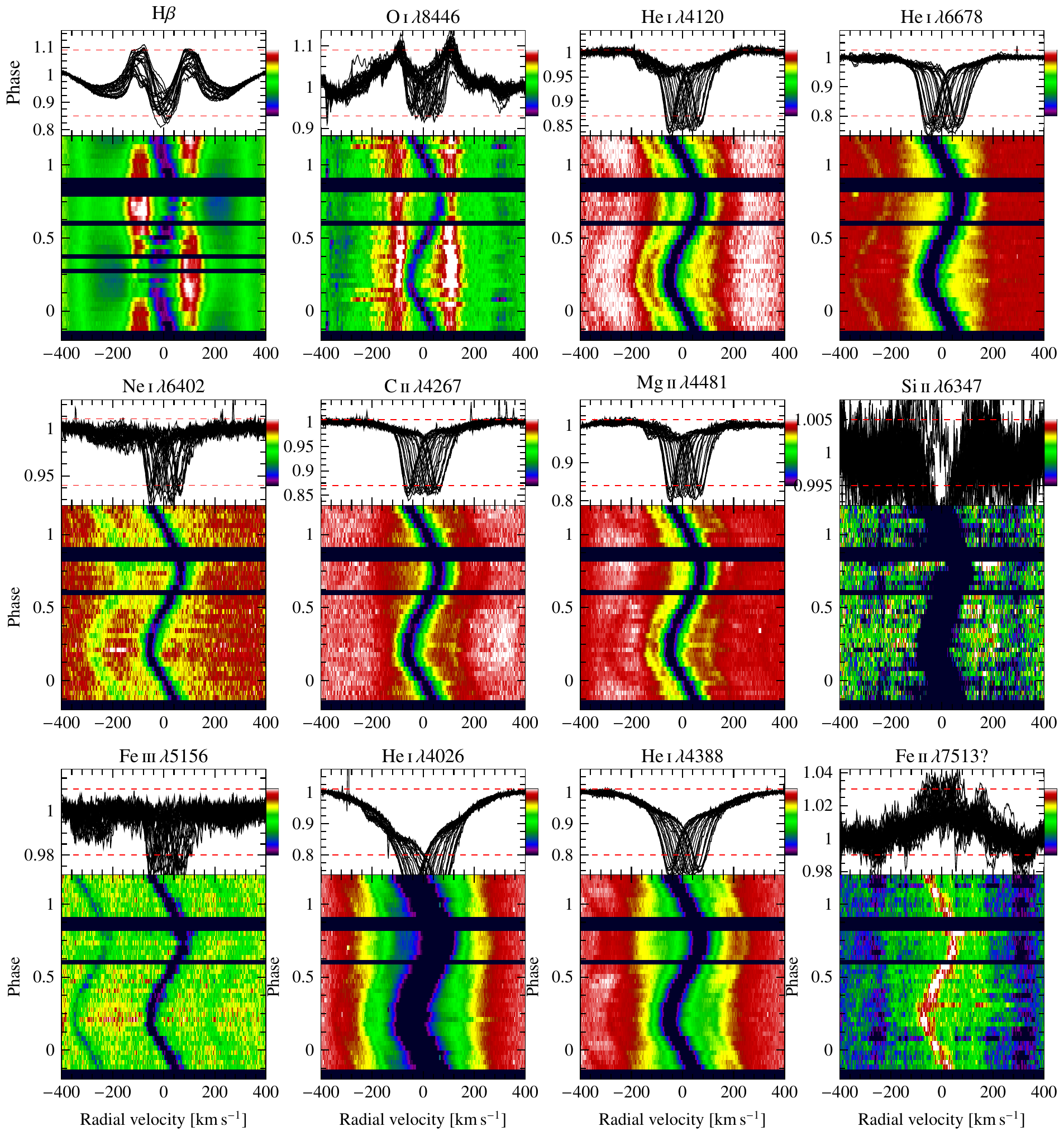}
   \caption{\label{appfig_HR6819_trails}
   Additional (see\ Fig.\,\ref{fig_HR6819_trails}) dynamical spectra of selected circumstellar and photospheric lines (as labeled) of the HR\,6819 system, phased with the orbital period (see Table~\ref{tab_comparison}) of the inner binary. The individual line profiles are overplotted at the top of each panel, where dashed horizontal lines identify the ranges of the color-coding in the dynamical spectra. H$\beta$\ was constructed with dataset B alone (see Table~\ref{tab_specobs}) to minimize the effect of the long-term disk variability of the Be star on the appearance of the emission. 
   }
\end{figure*}

The star that is part of the inner binary has been classified as B3\,III on account of its sharp lines \citep[see][]{1982ApJS...50...55S}.  The outer Be star stands out by its emission lines, but the shallowness of the broad photospheric absorption lines prevents a direct classification other than being similar to that of the inner B star. While in an unrecognized composite spectrum, the accuracy of spectral classifications can be compromised, in HR\,6819 there are good arguments that both temperature and luminosity of the inner B star are close to B3\,III. The spectral typing is relevant for this work because the minimum mass of the unseen object depends, through the mass function, on the minimum mass of its narrow-line companion in the inner binary.  When we consider the isolines in the Hertzsprung-Russell diagram, determining the minimum mass of a star requires determining at least a minimum temperature or a minimum luminosity, and preferably both.

For a minimum temperature, ratios of lines from the same element but different ionization stages can be good indicators. \textcolor{black}{Strong \ion{Fe}{ii} and \ion{Fe}{iii} lines of the narrow component} lie in the range between 510 and 520nm.  The \ion{Fe}{ii}$\lambda$\,5169 is stronger than \ion{Fe}{iii}$\lambda\lambda$\,5127,5156 by about a factor of two. For this line ratio, synthetic ATLAS9/SYNSPEC spectra \citep{2018A&A...609A.108S} show a monotonic growth of the temperature from 16\,kK for $\log g \sim 2.5$ to 18\,kK for $\log g \sim 3.5$ (see Fig.~\ref{appfig_ionbalance}). This places the narrow-line inner B star indeed in the B3 range \citep[e.g.,][assume 17100\,K as a typical $T_{\rm eff}$ for a B3\,III star in their calibration]{2010AN....331..349H}.

Similar arguments can be made for the luminosity class of the narrow-line component. As a first step, it is important to note that \ion{He}{i} lines are insensitive to temperature in the spectral range B2 to B3.  By contrast, they are sensitive to gravity, and hence luminosity, through the increasing pressure broadening as $\log g$ increases \citep{1998A&A...330..306L}. 
 
The \ion{He}{i} lines of the inner B3\,III star are less broadened than those of the Be star.  This is revealed by the difference in RV variability between the two stars:  The outer limits of the blends of the pressure-broadened \ion{He}{i} lines do not participate in the RV variation of the metal lines (see  \ion{He}{i}$\lambda\lambda$4026,4388 in  Fig.\,\ref{appfig_HR6819_trails}). Because the Be star does not exhibit any noticeable RV variations (as is evident from the quasi-stationary line emission), the also stationary outer wings of the \ion{He}{i} lines must be due to the Be star.  Because the stationary wings are wider, the pressure broadening of the Be star is higher.  By implication, the lower broadening of the inner B star indicates a lower gravity, that is, higher luminosity.  Accordingly, the inner B star is more evolved than the Be star.

An upper bound on the luminosity can be deduced from the presence or absence of emission lines that form deep in the photosphere \citep{2015IAUS..307..228R}. They become stronger as the photospheric density decreases, that is, the scale height and luminosity increase. In HR\,6819, such emission lines are present at $\lambda\lambda$ $\sim$\,6240\,\AA\ (a group of four lines), 7513\,\AA, and 7849\,\AA\  (see Fig.\,\ref{appfig_HR6819_trails} for an example). Candidate identifications of the two latter features are   \ion{Fe}{ii}$\lambda$7513.1767 and the \ion{Si}{ii}$\lambda\lambda$7848.80,7849.72 doublet. Some of these lines are also present in \object{BW\,Vul} and \object{$\xi^1$\,CMa, which have}  spectral types B2\,III and B0.7\,IV, respectively \citep{2015IAUS..307..228R}. \textcolor{black}{No other such emission lines} were found in HR\,6819. Because these emission lines become stronger and more abundant at higher luminosity, a luminosity class of IV--III is suggested for the inner B star, and that of the Be star is V. 

In conclusion, the spectrum of the inner narrow-line star agrees with a 16-18\,kK star toward the end of the main sequence. 
In the evolutionary tracks studied by \citet{2012A&A...537A.146E}, for instance, this means a star of at least 5\,\mbox{${\rm M}_{\odot}$}, and most likely of higher mass. Supportive evidence is available from \citet{2010AN....331..349H}, who listed masses for 56 B3\,III stars; none of them has a median value lower than 5\,\mbox{${\rm M}_{\odot}$}.  Therefore, 5\,\mbox{${\rm M}_{\odot}$}\ is a robust minimum mass for the inner B3\,III star that is suitable to establish a similarly robust minimum mass of the unseen companion.

\subsection{\textcolor{black}{Limits on the companion detectability}}
\label{app_maxmassdiff}
\textcolor{black}{The previous section considered the question of the minimum mass of the inner system and concluded that a second luminous star therein with the found minimum mass could not remain undetected. As is seen from the structure of the mass function (Appendix\ \ref{app_RV}), however, not only the minimum mass is important, but the mass difference between the B star and its companion matters as well.  This difference increases with increasing B star mass, and the associated larger difference in luminosity would make it easier for a luminous companion to hide.  The median mass values of \citet{2010AN....331..349H} used above for the determination of the upper mass limit for the B3\,III star {include two outliers} at $\text{about }10$\,\mbox{${\rm M}_{\odot}$}\ 
while all others are below 8\,\mbox{${\rm M}_{\odot}$}.  With an analogous reckoning to that used in Sect.\ \ref{sect_innerorbit} above, this suggests a minimum mass of the companion of the unseen companion of 5.6\,\mbox{${\rm M}_{\odot}$}.}

\textcolor{black}{\citet[][for the masses]{2010AN....331..349H} and \citet[][for the absolute magnitudes, see also Sect.\ \ref{sect_innerorbit} for an explanation]{2006MNRAS.371..185W} implied that for any combination of main-sequence masses between 5 and 4.2\,\mbox{${\rm M}_{\odot}$}\ (minimum masses of the two inner system companions as derived in Sect.\,\ref{sect_innerorbit}) and 8 and 5.6\,\mbox{${\rm M}_{\odot}$}\ (maximum masses, as derived above in this section), respectively, the magnitude difference $\Delta M_V$ does not exceed 1.7\,mag. It is larger for the lower set of masses. As we discuss below, it is not plausible that a companion with such a small magnitude difference could be missed in the available data, even for a critically rotating companion.}

\subsection{Parameters relative to similar B stars}
\label{app_bestfit}
The previous subsection aimed at deriving the minimum mass of the inner B3\,III star and, thereby, that of the unseen component. However, more realistic properties that are not based on skewed choices toward a minimum mass are needed for the use of the inner B star as the pivotal element for the analysis of all other properties, especially those of the Be star.  This is achieved best by comparison to calibrations from larger samples, or analyses of individual similar stars. The following will pursue both avenues.  Any systematic errors for the inner B3\,III star will propagate into similar systematic errors for the Be star, but will leave all differential conclusions intact.  Independent absolute analyses of the two stars would achieve this less easily. 

The calibration of B3\,III stars performed by \citet[][based on multicolor photometry, Hipparcos parallaxes, and different evolutionary models for 68 B3\,III stars]{2010AN....331..349H} gives a mean mass of $6.31\pm0.7$\,\mbox{${\rm M}_{\odot}$}\ with an assumed $T_{\rm eff}=17100$\,K  for B3\,III stars.  A broader range of physical parameters was derived for the  B3\,III star 18\,Peg by \citet[object 15 in their list]{2014A&A...566A...7N}. They found $5.8\pm0.4$ M$_{\odot}$  and also determined the temperature ($T_{\rm eff}=15.8\pm0.2$\,kK), gravity ($\log g=3.75\pm0.05$), radius ($5.5\pm0.5$\,\mbox{${\rm R}_{\odot}$}), and evolutionary age ($10^{7.8\pm0.1}$\,Myr). The SB1 nature of 18\,Peg (see below) was not known at the time.  However, from the absence of any spectroscopic signature of a possible secondary,  \citet{2016A&A...591L...6I} placed an upper limit of only 7\% on the flux contribution by a secondary at visual wavelengths.

In their dedicated study of 18\,Peg,  \citet{2016A&A...591L...6I} derived the parameter values photometrically, which in particular concerning gravity, is less reliable than modern spectroscopic methods.  Their rather low value of $\log g$ ($3.41\pm0.15$) is reflected in a surprisingly high radius ($8.4\pm1.6$\,\mbox{${\rm R}_{\odot}$}) and luminosity ($4000\pm1400$\,\mbox{${\rm L}_{\odot}$}). The calculated mass is $6.9\pm0.7$\,\mbox{${\rm M}_{\odot}$}.  It is noteworthy that according to the note to Table 1 of \citet{2016A&A...591L...6I}, these values were obtained using the evolutionary tracks of \citet{2012A&A...537A.146E}. However, as we show in the next section in Fig.~\ref{appfig_HRD}, they are only compatible with a star that is in the rapidly progressing expansion phase toward the giant branch.  The parameters derived in the same study by asteroseismology ($\log g=3.22\pm0.01$, $10.9\pm0.15$\,\mbox{${\rm R}_{\odot}$}) are even farther off, and \citet{2016A&A...591L...6I} pointed out that the stated uncertainties only reflect the statistical part, not any potential systematic offsets. Adoption of the evolutionary tracks by \citet{2011A&A...530A.115B} might lead to a more plausible star.  However, considering that 18\,Peg is only used as a proxy for a typical B3\,III star and the parameter values from \citet{2014A&A...566A...7N} agree reasonably well with the calibration by \citet{2010AN....331..349H}, these values appear adequate for the purpose of characterizing the B3\,III star in HR\,6819.  Therefore, while 5\,\mbox{${\rm M}_{\odot}$}\ is the absolute minimum mass of the inner B star (Appendix\,\ref{app_minmass}), the range of 6 to 7\,\mbox{${\rm M}_{\odot}$}\ is more suitable for all other purposes.

\subsection{Distance to the system}
\label{subapp_distance}
With these parameters, distance can be addressed. Neither Gaia nor Hipparcos parallaxes take binarity into account, and therefore do not reach their nominal accuracy  (Sect.\,\ref{sect_trajectory}). An alternative distance estimate can be made with UV fluxes measured with the {\it International Ultraviolet Explorer} (IUE) satellite \citep{1978Natur.275..372B}\footnote{Data available from the Mikulski Archive for Space Telescopes: \url{https://dx.doi.org/10.17909/t9-n3zj-zv42}}. IUE obtained two high-resolution spectra, one in the long- and one in the short-wavelength range. Both observations used the large aperture so that they are suitable to determine absolute fluxes.  The spectra were resampled in a flux-conserving way with a 2\,\AA\ bin to reduce the noise. For the dereddening of the measured fluxes, the reddening $E(B-V)=0.135$\,mag listed by \citet{2002BaltA..11....1W} for HR\,6819 and the extinction parameterization by \citet{1989ApJ...345..245C} were applied together with the standard value of the Galactic total-to-selective extinction ratio, $R_V=3.1$.  

\begin{figure}[t]
   \centering
   \includegraphics[width=0.5\textwidth]{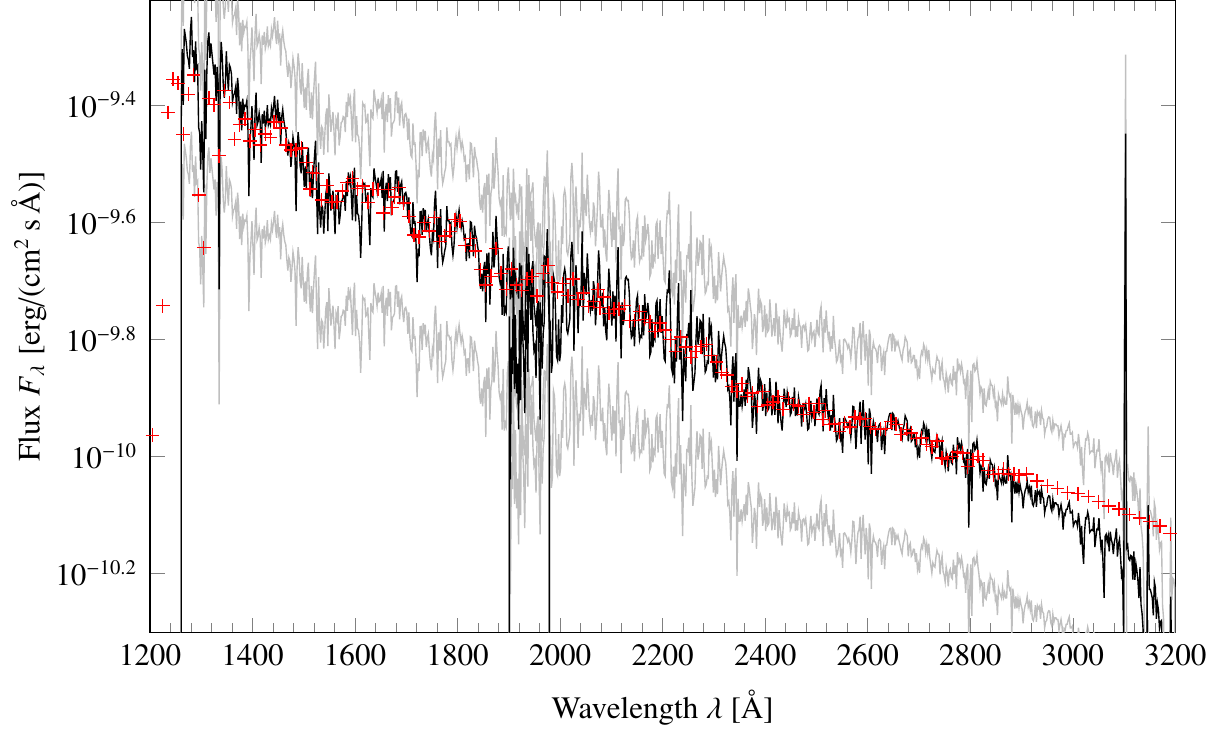}
   \caption{\label{appfig_IUE}  Binned and dereddened IUE spectra (black) of HR\,6819.  They are scaled to 55\% to represent the flux contribution from the inner  B3\,III component to the total flux; spectra plotted in gray indicate 35-75\% ranges (Sect.\,\ref{subapp_distance}.) The theoretical SED for $T_{\rm eff}=16$\,kK and $\log g=3.5$ \citep{2003IAUS..210P.A20C} is overplotted in red, multiplied by $6.2\times10^{-18}$ to account for distance and stellar radius. }
\end{figure}

Model fluxes corresponding to the parameters estimated above are available from \citet[it should be noted that these were adapted to $F_\lambda$ surface flux units]{2003IAUS..210P.A20C}\footnote{Downloadable from: \url{http://www.stsci.edu/hst/instrumentation/reference-data-for-calibration-and-tools/astronomical-catalogs/castelli-and-kurucz-atlas}}, although only for $T_{\rm eff}=16$\,kK and $\log g=3.5$ and $4.0$. For the comparison here, $\log g=3.5$ was chosen, but the difference to $4.0$ is marginal in view of the other uncertainties discussed below.

Because HR\,6819 is a multiple star, the flux contributions by each component have to be taken into account. Preliminary results from \textcolor{black}{disentangling}
the spectra indicate that the inner B3\,III and the outer Be star are about equally bright in the visual regime and that the Be star has a slightly higher temperature and a somewhat higher $\log g$ (Hadrava et al., in prep).  The similarity in flux is corroborated by the equivalent widths (EWs) of \ion{He}{i} lines of the two stars when lines with only a small contribution from pressure broadening are chosen.  Across the given temperature and luminosity domain, they are fairly constant, so that the EW ratio is a good proxy of the luminosity ratio \citep{1998A&A...330..306L}.  For instance,  the total EW of \ion{He}{i}$\lambda4713$ is about 0.3\,\AA, to which the narrow component alone contributes 50 to 60\%. That is, the two stars are about equally bright, and a contribution to the total IUE flux by the B3\,III component of $55\pm20$\% seems a reasonable estimate.  

Figure\,\ref{appfig_IUE} depicts the observed absolute UV flux distribution of the HR\,6819 system as characterized above and scaled to 55\% for comparison to the theoretical Castelli \& Kurucz model of the inner B3\,III star alone.  The slope indicates a slightly hotter overall flux distribution (i.e., the sum of the modeled inner B3\,III and the outer Be star), which agrees with the preliminary  \textcolor{black}{disentangling}
solution of a slightly hotter outer Be star.  The match of observed IUE flux and theoretical spectral energy distribution (SED) can also be appreciated in Fig.~\ref{appfig_IUE}.

For the flux definition of the  \citet{2003IAUS..210P.A20C} grid as offered by the STScI (see above), distances $D$ are simply obtained as $D = R\sqrt{f_{\rm s}}$, where $R$ is the stellar radius and $f_{\rm s}$ is the distance-dependent dimensionless flux-scaling factor, which for the  B3\,III star HR\,6819 is $f_{\rm s}=6.2\pm1\times10^{18}$ (see Fig.\,\ref{appfig_IUE}).
\textcolor{black}{Considering that the uncertainties of the result are currently fully dominated by other factors such as the flux ratio of the components  (as seen by the range between the gray SEDs indicated in Fig.\ \ref{appfig_IUE}), a simple visual approximation was sufficiently precise to derive $f_{\rm s}$.
}
Adopting $5.5\pm0.5$\,\mbox{${\rm R}_{\odot}$}\ as discussed above results in a distance to the system of $310\pm60$\,pc. The uncertainty is dominated by the loosely constrained flux distribution between the two luminous components in the system, with the second largest contribution coming from the stellar radius. Overall, the larger the relative flux contribution by the  B3\,III component, the more nearby the system, while the distance would increase with the B3\,III radius. 

Alternative plausible values for temperature, radius, reddening, and selective absorption as well as fitting the observed SED with two theoretical SEDs with flux ratios in the above range (35--75\%) alter the distance and its uncertainty only marginally:  While the distance changes by less than a dozen parsec, the uncertainty range is always several dozens of parsec.  This uncertainty is again dominated by the flux ratio between the two stars.  Only with a better constrained flux ratio can a more precise distance value be derived from the SED.

\section{Properties of the outer Be star and the age of the system} \label{app_Bestar}
 
\begin{figure}[t]
   \centering
   \includegraphics[width=0.5\textwidth]{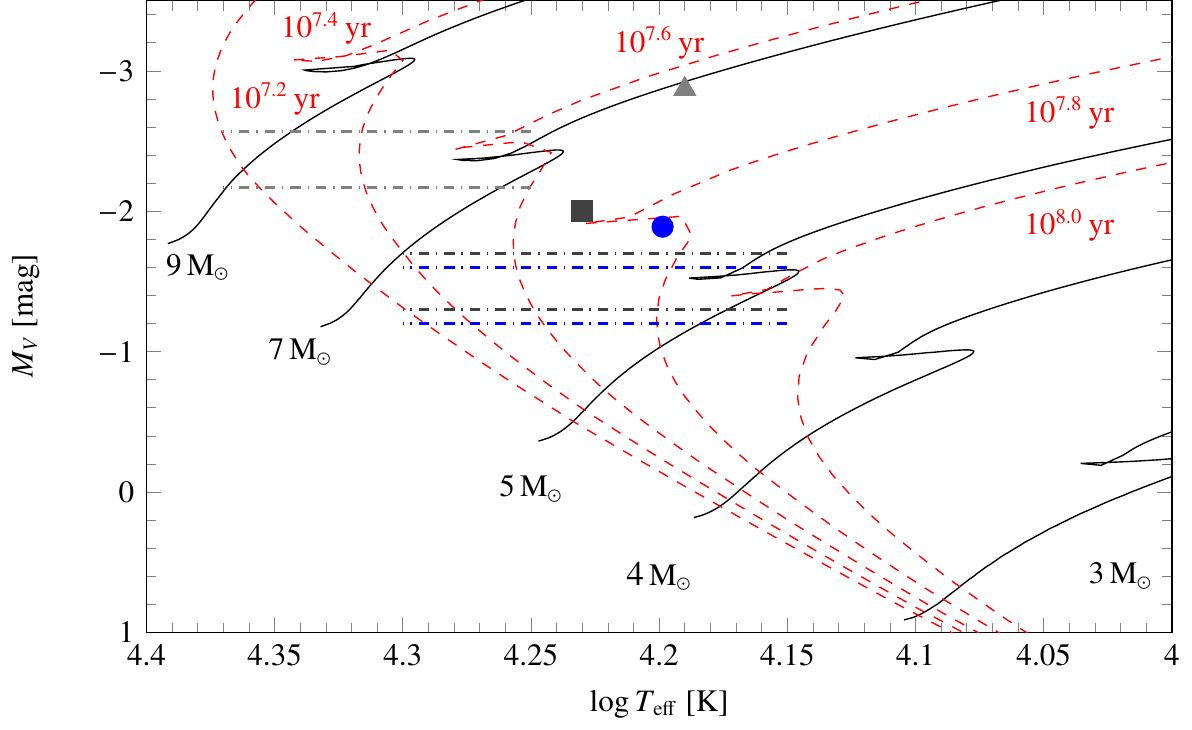}
   \caption{\label{appfig_HRD}Evolutionary tracks (solid) and isochrones (dashed) from \citet{2012A&A...537A.146E}. 
Three potential locations of B3\,III stars are marked by filled shapes: 18\,Peg as modeled by \citet[blue filled circle]{2014A&A...566A...7N} as well as by \citet[gray triangle]{2016A&A...591L...6I}, and the generic calibration by \citet[dark gray square]{2010AN....331..349H}, 
all discussed in Appendix~\ref{app_distance}. The horizontal dash-dotted lines give the respective implied ranges of the intrinsic brightness of the Be star in HR\,6819.  The blue and dark gray lines assume that the apparent magnitude of the Be star is the same as that of a B3\,III star as determined by \cite{2014A&A...566A...7N} and \citet{2010AN....331..349H}, respectively. Each pair of lines with the same color indicates a disk excess $\Delta V$ of $-0.3$ and $-0.7$\,mag, respectively. 
}
\end{figure}
 
 As illustrated by Fig.~\ref{fig_HR6819_trails} and preliminarily quantified in Appendix~\ref{subapp_distance} through the EW of \ion{He}{i} 4713, the outer Be star is of similar spectral type and similar visual magnitude as the inner B3\,III star. This is also supported by the  \textcolor{black}{disentangling of spectra}
(Hadrava et al., in prep.).  
 However, this similarity does not imply that the Be star itself shares the other parameter values of the inner star as well because the observed flux is the sum of the stellar and a circumstellar component.  There is a well-known magnitude excess in Be stars that is caused by scattering of stellar light in the circumstellar disk, of up to $\Delta V\sim-0.5$\,mag \citep{2012ApJ...756..156H}. Because the Be star in HR\,6819 has a strong \mbox{H$\alpha$}\ emission and low inclination, as follows from the shape of the emission \citep[see][]{2000A&A...359.1075H, 2013A&ARv..21...69R}, this upper limit is adopted with a conservative range of $-0.3$ to $-0.7$\,mag.   Rapid rotation is another mechanism that may alter the apparent parameters of a star in an aspect-angle-dependent way \citep{2013A&ARv..21...69R}.  However, the effect is too small in comparison to other uncertainties to be considered here.
 
 Figure~\ref{appfig_HRD} identifies the potential location of the Be star in HR\,6819 (i.e., after correction for the circumstellar flux excess) relative to a B3\,III star in a theoretical Hertzsprung-Russell diagram (HRD), along with evolutionary tracks and isochrones. The underlying values correspond to those derived for 18\,Peg by \citet{2013A&A...550A..26N} and by \citet{{2016A&A...591L...6I}} and to the general calibration by \citet{2010AN....331..349H}, see also Appendix\,\ref{app_distance}. The choice of 18\,Peg does not pretend to present the true values for HR\,6819, but merely serves as an anchor for the discussion of the values of temperature, luminosity, and age of the Be star in HR\,6819 relative to the inner B3\,III star. The luminosities, $L$, were converted into absolute visual magnitudes, $M_{\rm V}$, using Eq.~10 of \citet{2013A&A...550A..26N} for the bolometric correction, so that for a single star with temperature $T_{\rm eff}$
\begin{equation}
M_V = -2.5 \log L +4.74 - ( 21-5.34\log T_{\rm eff}).
\end{equation}
No correction is included for binarity, and the calculated $M_V$ is independent of reddening, which has to be accounted for when the apparent magnitude is derived from the absolute magnitude. 
In the following discussion, the properties of the Be star refer to the Be star alone (without contributions from the circumstellar disk to the brightness).  

The evolutionary age of B3\,III stars is about $65\pm15$\,Myr (see the result for 18\,Peg by \citealp{2014A&A...566A...7N}, which is based on computations by \citealp{2012A&A...537A.146E}).  When similar values as by \citealp{2014A&A...566A...7N} are assumed and both stars in HR\,6819 evolved unaffected by binary interaction, the isochrones in Fig.\,\ref{appfig_HRD} (dark gray symbols) suggest a common age of about 70\,Myr. However, when instead parameter values such as those by \citet{2016A&A...591L...6I} are assumed, the individual evolutionary ages of the two stars are less easily reconciled under the assumption of 
\textcolor{black}{single star evolution for both stars}, namely with an age of about 40\,Myr, but the Be star would have to be at the very end of its core hydrogen-burning phase, and the B star well beyond it (see Fig.~\ref{appfig_HRD}, light gray symbols).  
 
The so-called Be phenomenon \citep{2013A&ARv..21...69R} is limited to the main sequence 
(taken as the core hydrogen-burning phase, which for B stars, includes luminosity classes V to III). After core-hydrogen exhaustion, expansion of the envelope causes the surface rotation to slow down far below the near-critical values that are needed for a Be star to form and sustain a Keplerian disk.  
With parameter values similar to those obtained by
\citet{2014A&A...566A...7N}, a B3\,III star is about to reach the end of its main sequence life. It follows that a star of equal age cannot have both similar $M_V$ and significantly lower T$_{\rm eff}$ or later spectral type (Fig.~\ref{appfig_HRD}) because it would fall well above the main sequence. Preliminary \textcolor{black}{disentangling}
of the spectra, as well as the relative line strengths over the full instrumental wavelength range, indeed indicate that the Be star may be slightly hotter than the B3\,III star and also have a slightly higher surface gravity. Fig.~\ref{appfig_HRD} therefore shows that the Be star must be on the same or on an earlier isochrone, and the B3\,III must be at least as old as the Be star.  Only if there had been significant binary interaction in the inner system could the B3\,III appear younger than the Be star because it would have acquired additional fuel and lifetime through Roche-lobe overflow from the BH progenitor.  

The current orbital separation of the inner binary makes it likely that the two stars have been interacting.   If the evolutionary age of the B3\,III star nevertheless remained unaltered or even increased relative to the Be star, more remote explanations need to be considered.  For instance, it appears conceivable that hydrogen-depleted material from the BH progenitor was transferred to the core of the inner B star, thereby reducing the operating time of the inner B star. Alternatively, if the Be star is itself a binary, it might have been rejuvenated as well and by more than the B3\,III star. This is not completely implausible: If Be stars owe their rapid rotation to spin-up by mass overflow from a now underluminous companion \citep[e.g.,][and references therein]{2020A&A...633A..40L} a resulting rejuvenation of the outer Be star is one possible explanation. Some blue stragglers in open clusters have been reported to show H$\alpha$ in emission and may be classical Be stars \citep{2018A&A...610A..30A, 2020A&A...634A..51B}.

When we disregard the possible difference between the age estimates, the outer Be star considered alone might be as young as $\sim 10^{7.2} = 15$\,Myr if the preliminary "slightly hotter and slightly higher gravity" result from the  \textcolor{black}{disentangling}
is taken at face value. A conclusive statement on this problem has to await the results of the full \textcolor{black}{disentangling}
of the spectra of the two stars (Hadrava et al., in prep.).  Under the current premises, the age bracket of 15 to 75\,Myr for the HR\,6819 system is the most plausible scenario, with a preference for the higher value.

\end{document}